\documentstyle[12pt]{article}
\begin{document}
\newcommand {\ber} {\begin{eqnarray*}}
\newcommand {\eer} {\end{eqnarray*}}
\newcommand {\bea} {\begin{eqnarray}}
\newcommand {\eea} {\end{eqnarray}}
\newcommand {\beq} {\begin{equation}}
\newcommand {\eeq} {\end{equation}}
\newcommand {\state} [1] {\mid \! \! {#1} \rangleg}
\newcommand {\sym} {$SY\! M_2\ $}
\newcommand {\eqref} [1] {(\ref {#1})}
\newcommand{\preprint}[1]{\begin{table}[t] 
           \begin{flushright}               
           \begin{large}{#1}\end{large}     
           \end{flushright}                 
           \end{table}}                     
\def\Acknowledgements{\bigskip  \bigskip {\begin{center} \begin{large}
             \bf ACKNOWLEDGMENTS \end{large}\end{center}}}

\newcommand{\half} {{1\over {\sqrt2}}}
\newcommand{\dx} {\partial _1}

\def\Dslash{\not{\hbox{\kern-4pt $D$}}}
\def\cmp#1{{\it Comm. Math. Phys.} {\bf #1}}
\def\cqg#1{{\it Class. Quantum Grav.} {\bf #1}}
\def\pl#1{{\it Phys. Lett.} {\bf #1B}}
\def\prl#1{{\it Phys. Rev. Lett.} {\bf #1}}
\def\prd#1{{\it Phys. Rev.} {\bf D#1}}
\def\prr#1{{\it Phys. Rev.} {\bf #1}}
\def\pr#1{{\it Phys. Rept.} {\bf #1}}
\def\np#1{{\it Nucl. Phys.} {\bf B#1}}
\def\ncim#1{{\it Nuovo Cimento} {\bf #1}}
\def\lnc#1{{\it Lett. Nuovo Cim.} {\bf #1}}
\def\jmath#1{{\it J. Math. Phys.} {\bf #1}}
\def\mpl#1{{\it Mod. Phys. Lett.}{\bf A#1}}
\def\jmp#1{{\it J. Mod. Phys.}{\bf A#1}}
\def\aop#1{{\it Ann. Phys.} {\bf #1}}
\def\mycomm#1{\hfill\break{\tt #1}\hfill\break}

\begin{titlepage}
\rightline{TAUP-2453-97}
\rightline{WIS-97/29/SEP-PH}
\rightline{\today}
\vskip 1cm
\centerline{{\Large \bf The string tension in massive $QCD_2$}}
\vskip 1cm
\centerline{A. Armoni$^a$
\footnote{e-mail: armoni@post.tau.ac.il}
, Y.Frishman$^b$
\footnote{e-mail: fnfrishm@wicc.weizmann.ac.il}
 and J. Sonnenschein$^a$
\footnote{e-mail: cobi@post.tau.ac.il \\
Work supported in part by the Israel Science Foundation,
and  the US-Israel Binational
Science Foundation }}
\vskip 1cm
\begin{center}
\em $^a$School of Physics and Astronomy
\\Beverly and Raymond Sackler Faculty of Exact Sciences
\\Tel Aviv University, Ramat Aviv, 69978, Israel
\\and
\\$^b$Department of Particle Physics
\\Weizmann Institute of Science
\\76100 Rehovot, Israel
\end{center}
\vskip 1cm

\begin{abstract}
We compute the string tension in massive $QCD_2$.
It is shown that the string tension vanishes when the
mass of the dynamical quark is zero, with no dependence on the
representations of the dynamical or of the external charges. When a small
mass  ($m\ll e$) is added, a tension appears and we calculate its value as a
function of the representations.
\end{abstract}
\end{titlepage}

 In a recent paper by Gross et al.\cite{gross} it was argued that
 two dimensional QCD exhibits a screening nature when the dynamical
quarks have no mass. Confinement appears when mass is given to the
quarks.  Solutions to the equations of motion and a large $N_f$
analysis which support this statement were given in refs.\cite{AFS}.

 Similar phenomena occur in two dimensional QED. It is well
known\cite{CJS} that integer charges can screen fractional charges
when the dynamical electrons are massless. The naive picture of
confinement is restored when a mass is given to the dynamical
electrons and when the external charge cannot be composed of the dynamical
charge.
The expression for the string tension in the abelian case is
\beq
\sigma = m\mu \left (1-\cos (2\pi {q_{ext} \over q_{dyn}}) \right ) \label{abelian},
\eeq
where $m$ is the electron mass, $\mu = e{\exp (\gamma) \over 2\pi^{3/2}}$
($e$ is the coupling and $\gamma$ is the Euler number)
 and $q_{ext}$,$q_{dyn}$ are the external and dynamical charges
respectively. 

In this note we generalize the proof of \cite{CJS} to the non-abelian
case. We show that when the dynamical quarks are massless, the external
source can be extracted from the action by a redefinition of the
(bosonized) matter field and therefore there is no string tension.
 When a mass term does exist, we use this redefinition
to calculate the string tension. The proof is given in the interesting
case of dynamical and external charges in the fundamental/adjoint
representations. Additional remarks concerning the symmetric and
anti-symmetric cases are also given.

The action of bosonized $QCD_2$ with massive quarks in the fundamental
representation of $SU(N)$ is the following \cite{FS}
\bea
\lefteqn{S_{fundamental}={1\over{8\pi}}\int _\Sigma d^2x \ tr(\partial _\mu
g\partial ^\mu g^\dagger) + } \label{fund} \\
 && {1\over{12\pi}}\int _B d^3y \epsilon ^{ijk} \
 tr(g^\dagger\partial _i g) (g^\dagger\partial _j g)(g^\dagger\partial _k g) +
 \nonumber \\
&& {1\over 2} m \mu _{fund} \int d^2x \ tr (g+g^\dagger)  
-\int d^2 x{1\over {4e^2}} F^a _{\mu \nu} F^{a \mu \nu} + \nonumber \\
&& -{1\over 2\pi}\int d^2 x \ tr (ig^\dagger \partial_+ g A_-
+ig\partial_ - g^\dagger A_+ + A_+ g A_- g^\dagger - A_+ A_-), \nonumber
\eea
where $e$ is the gauge coupling, $m$ is the quark mass, $\mu \sim e$,
 $g$ is an $N\times N$ unitary matrix, $A_\mu$ is the gauge field and the
trace is over $U(N)$ indices. Note, however, that only
 the $SU(N)$ part of the matter field $g$ is gauged.

When the quarks transform in the adjoint representation, the
expression for the action is\cite{AGSY}\footnote{Since we do not consider
  instantons, the remark of ref.\cite{smilga} about the bosonized form
  of gauged adjoint matter is not crucial to our discussion.} 
\bea
\lefteqn{S_{adjoint}={1\over{16\pi}}\int _\Sigma d^2x \ tr(\partial _\mu
g\partial ^\mu g^\dagger) + } \label{adjoint} \\
 && {1\over{24\pi}}\int _B d^3y \epsilon ^{ijk} \
 tr(g^\dagger\partial _i g) (g^\dagger\partial _j g)(g^\dagger\partial
 _k g) +
 \nonumber \\
&& {1\over 2} m \mu _{adj} \int d^2x \ tr (g+g^\dagger)  
-\int d^2 x{1\over {4e^2}} F^a _{\mu \nu} F^{a \mu \nu} + \nonumber \\
&& -{1\over 4\pi}\int d^2 x \ tr (ig^\dagger \partial_+ g A_-
+ig\partial_ - g^\dagger A_+ + A_+ g A_- g^\dagger - A_+ A_-) \nonumber
\eea
It differs from \eqref{fund} by a factor of one half in front of the
 WZW and interaction terms because $g$ is real and represents Majorana
fermions. Another difference is that $g$ now is an
$(N^2-1)\times(N^2-1)$ orthogonal matrix. The two actions 
\eqref{fund} and \eqref{adjoint} can be schematically
 represented by the following action 
\beq
S=S_0 + {1\over 2} m \mu _R \int d^2x \ tr (g+g^\dagger) 
 -{i k_{dyn}\over 4\pi}\int d^2 x \ (g\partial_ - g^\dagger)^a A_+^a, \label{universal}
\eeq
 where $A_- = 0$ gauge was used, $S_0$ stands for the WZW action and
 the kinetic
 action of the gauge field. $k_{dyn}$ is the level (the chiral
 anomaly) of the dynamical
 charges ($k=1$ for the fundamental representation of $SU(N)$ and $k=N$
 for the adjoint representation).

Let us add an external charge to the action. We choose static
(with respect to the light-cone coordinate $x^+$) charge and
therefore we can omit its kinetic term
from the action. Thus an external charge coupled to the gauge field
would be represented by
\beq
-{i k_{ext}\over 4\pi}\int d^2 x \ (u\partial_ - u^\dagger)^a A_+^a
\eeq
Suppose that we want to put a quark and an anti-quark at a very large
separation. A convenient choice of the charges would be a
direction in the algebra in which the generator has a diagonal
form. The simplest choice is a generator of an $SU(2)$ subalgebra. As an
example we write down the generator in the case of fundamental and
adjoint representations.
\bea
 && T^3_{fundamental} = diag ({1\over 2},-{1\over 2},\underbrace{0,0,...,0}_{N-2}) \\
 && T^3_{adjoint} = diag (1,0,-1,
\underbrace{{1\over 2},-{1\over 2},{1\over 2},-{1\over 2},...,{1\over
    2},-{1\over 2}}_{2(N-2)\,\,\, doublets},\underbrace{0,0,...,0}_{(N-2)^2})
\eea  
Generally $T^3$ can be written as
\beq
 T^3 = diag (\lambda _1,\lambda _2,...,\lambda
 _i,0,0,...),
\eeq
where $\{ \lambda _i \}$ are the 'isospin'
 components of the representation under the
$SU(2)$ subgroup.

We take the $SU(N)$ part of $u$ as (see Appendix)\footnote{In the case
  of $SU(2)$ the choice is $u=\exp -i2\pi \left ( \theta (x^-+L )-\theta
  (x^--L )\right ) T^3_{ext}$} 
\beq
\label{external-charge}
u=\exp -i4\pi \left ( \theta (x^-+L )-\theta
  (x^--L )\right ) T^3_{ext},
\eeq
where $T^3_{ext}$ represents the '3' generator of the external charge and
$u$ is static with respect to the light-cone time coordinate $x^+$.
The theta function is used as a limit of a smooth
function which interpolates between 0 and 1 in a very short
distance. In that limit $u=1$ everywhere except at isolated points,
where it is not well defined.  

The form of the action \eqref{universal} in the presence of an external
source is 
\bea
\lefteqn{
S=S_0 + {1\over 2} m \mu _R \int d^2x \left \{ \ tr
  (g+g^\dagger) +\right . } \\
&&
  \left . [ -{i k_{dyn}\over 4\pi} (g\partial_ - g^\dagger)^a
   +  k_{ext} \delta^{a3} ( \delta (x^-+L)-\delta
   (x^--L))] A_+^a \right \} \nonumber
\eea
The external charge can be eliminated from the action by a
transformation of the matter field. A new field $\tilde g$ can be
defined as follows
\bea
\lefteqn{
-{i k_{dyn}\over 4\pi} (\tilde g\partial_ - \tilde
g^\dagger)^a=} \\
&&
-{i k_{dyn}\over 4\pi} (g\partial_ - g^\dagger)^a 
+k_{ext} \delta^{a3}\left ( \delta (x^-+L)-\delta
   (x^--L) \right )   \nonumber
\eea
This definition leads to the following equation for $\tilde g^\dagger$
\beq 
\partial _- \tilde g^\dagger = \tilde g^\dagger \left (g\partial _-
  g^\dagger
+i4\pi {k_{ext} \over k_{dyn}} (\delta (x^-+L)-\delta
   (x^--L) ) T^3_{dyn} \right ) \label{tilde}
\eeq
 The solution of \eqref{tilde} is 
\bea 
\lefteqn{
 \tilde g^\dagger =} \\
&&  P \exp \int dx^- \left ( g\partial _-  g^\dagger
 +i4\pi {k_{ext} \over k_{dyn}} (\delta (x^-+L)-\delta
   (x^--L) ) T^3_{dyn} \right ) = \nonumber \\
 &&  e^{i4\pi {k_{ext} \over k_{dyn}}\theta(x^-+L) T^3_{dyn}}
 g^\dagger
e^{-i4\pi {k_{ext} \over k_{dyn}}\theta(x^--L) T^3_{dyn}} , \nonumber
\eea
where $P$ denotes path ordering and we assumed that $T^3_{dyn}$
commutes with $g\partial _- g^\dagger$ for $x^- \ge L$ and with $g^\dagger$
for $x^-=-L$ (as we shall see, this assumption is self consistent
with the vacuum configuration).

 Let us take the limit $L\rightarrow \infty$. For $-L<x^-<L$, the above relation simply means that
\beq 
g=\tilde g  e^{i4\pi {k_{ext} \over k_{dyn}} T^3_{dyn}} \ 
\eeq

Since the Haar
measure is invariant (and finite, unlike the fermionic case) with respect to unitary transformations,
the form of the action in terms of the new variable $\tilde g$ reads
\bea
\lefteqn{
S=S_{WZW}(\tilde g) + S_{kinetic}(A_{\mu}) 
-{i k_{dyn}\over 4\pi}\int d^2 x \ (\tilde g\partial_ - \tilde
g^\dagger)^a A_+^a }
\label{rotate} \\
&&
+ {1\over 2} m \mu _R \int d^2x \
 tr (\tilde g  e^{i4\pi {k_{ext}  \over k_{dyn}} T^3_{dyn}}   
+ e^{-i4\pi {k_{ext}  \over k_{dyn}} T^3_{dyn}} \tilde g^\dagger ) 
  \nonumber
\eea
which is $QCD_2$ with a chiraly rotated mass term.

The string tension can be calculated easily from \eqref{rotate} \cite{CJS}. It is
simply the vacuum expectation value of the Hamiltonian density,
relative to
the v.e.v. of the Hamiltonian density of the theory without an external source,
\beq
\sigma = <H>-<H_0>
\eeq
The vacuum of the theory is given by $\tilde g =1$.
\bea
\lefteqn{
<H>= } \\
&&
-  {1\over 2} m \mu _R \
 tr ( e^{i4\pi {k_{ext}  \over k_{dyn}} T^3_{dyn}}   
+ e^{-i4\pi {k_{ext}  \over k_{dyn}} T^3_{dyn}} )
= \nonumber \\
&&
 -   m \mu _R \
\sum _i \cos (4\pi \lambda _i  {k_{ext}  \over k_{dyn}} ) \nonumber
\eea 
Therefore the string tension is
\beq
\sigma =  m \mu _R \
\sum _i \left ( 1-\cos (4\pi \lambda _i  {k_{ext}  \over k_{dyn}}
  )\right ) \label{sigma}
\eeq

A few remarks should be made:

(i) The string tension \eqref{sigma} reduces to the abelian string tension
\eqref{abelian} when abelian charges are considered. It seems that the
non-abelian generalization is realized by replacing the charge $q$ with
$k$.

(ii) The string tension was calculated in the tree level of the
bosonized action. Perturbation theory (with $m$ as the
coupling) may change eq.\eqref{sigma}, but we believe that it would
not change its general character.
 
(iii) When no dynamical mass is present, the theory exhibits screening. This
is simply because non-abelian charges at the end of the world interval
 can be eliminated from the action by
a chiral transformation of the matter field.

(iv) When the test charges are in the adjoint representation
$k_{ext} = N$,  equation \eqref{sigma} predicts screening by the fundamental
charges (with $k_{dyn} =1$).

(v) String tension appears when the test charges are in the fundamental
representation and the dynamical charges are in the adjoint
\cite{witten}. The value of the string tension is
\beq
\sigma = m \mu _{adj} \left (
2(1-\cos {4\pi \over N}) + 4(N-2)(1-\cos {2\pi \over N}) \right)
\eeq

 The case of $SU(2)$ is special. The $4\pi$ which
  appears in eq.\eqref{sigma} is replaced by $2\pi$, since the
 bosonized form of the external $SU(2)$ fundamental matter differs by
 a factor of a half with respect to the other $SU(N)$ cases
 (see Appendix). Hence, the string tension in this case is $4m \mu _{adj}$.

The generalization of \eqref{sigma} to arbitrary representations is
not straightforward. However, we can comment about its nature (without
rigorous proof).

Let us focus on the interesting case of the antisymmetric
representation. One can show that in a similar manner to \cite{AGSY},
the WZW action with $g$ taken to be ${1\over 2}N(N-1) \times {1\over
  2}N(N-1)$ unitary matrices, is a bosonized version of $QCD_2$ with
fermions in the antisymmetric representation. 

The antisymmetric representation is described in the
Young-tableaux notation
 by two vertical boxes. Its dimension is ${1\over 2}N(N-1)$ and its diagonal
$SU(2)$ generator is
\bea
T^3_{anti-symmetric} = diag (
\underbrace{{1\over 2},-{1\over 2},{1\over 2},-{1\over 2},...,{1\over
    2},-{1\over 2}}_{(N-2)\,\,\, doublets},0,0,...,0),
\eea
 and consequently $k= N-2$. When the dynamical charges are
 in the fundamental and the external in the antisymmetric the string
 tension should vanish because tensor product of two fundamentals
 include the antisymmetric representation. Indeed,
 \eqref{sigma} predicts this result.

The more interesting case is when the dynamical charges are
antisymmetric and the external are fundamentals. In this case the
value of the string tension depends on whether $N$ is odd or even\cite{witten}.
When $N$ is odd the string tension should vanish because the
anti-fundamental
representation can be built by tensoring the antisymmetric
representation with itself ${1\over 2}(N-1)$ times. When $N$ is even
 string tension must exist. Note that \eqref{sigma} predicts
\bea
\label{as1}
\sigma = 2m\mu _{as} (N-2)(1-\cos {2\pi \over N-2})
\eea
which is not zero when $N$ is odd.

The resolution of the puzzle seems to be the following.
Non-Abelian charge can be static with respect to its spatial
location. However, its representation may change in time due to
emission or absorption of soft gluons (without cost of energy). 
Our semi-classical
description of the external charge as a c-number is insensitive to
this scenario. We need an extension of 
\eqref{external-charge} which takes into account the possibilities of
all various representations. We propose the following external current  
\beq
\label{external-current}
j^a _{ext}=\delta ^{a3} k_{ext}(1+lN)(\delta(x^-+L)-\delta(x^--L))
\eeq
where $l$ is an arbitrary positive integer. The value of the external current corresponds to
$1+lN$ charges which were multiplied in a symmetric way.   
The resulting string tension is 
\beq
\sigma =  m \mu _R \
\sum _i \left ( 1-\cos (4\pi \lambda _i  {k_{ext}  \over k_{dyn}}(1+lN)
  )\right ),
\eeq
which includes the arbitrary integer $l$. What is the value of $l$
that we should pick ?

The dynamical charges are attracted to the external charges in such a way
that the total energy of the configuration is minimal. Therefore the
value of $l$ which is needed, is the one that guarantees minimal string
tension.

Thus the extended expression for string tension is the following 
\beq
\sigma =  \min _k \left \{ m \mu _R \
\sum _i \left ( 1-\cos (4\pi \lambda _i  {k_{ext}  \over k_{dyn}}(1+lN)
  )\right ) \right \}\label{sigma2}
\eeq

In the case of dynamical antisymmetric charges and external
fundamentals and odd $N$, $l={1\over 2}(N-3)$ gives zero string
tension. When $N$ is even the string tension is given by \eqref{as1}.

The expression \eqref{sigma2} yields the right answer in some
other cases also.

Another example is the case of
dynamical charges in the symmetric representation. The bosonization
for this case can be derived in a similar way to that of the
antisymmetric representation, and $T^3$
is given by
\beq
T^3_{symmetric} = diag (1,0,-1,
\underbrace{{1\over 2},-{1\over 2},{1\over 2},-{1\over 2},...,{1\over
    2},-{1\over 2}}_{(N-2)\,\,\, doublets},0,0,...,0),
\eeq
 and therefore $k=N+2$. When the external charges transform in the fundamental
representation and $N$ is odd, eq.\eqref{sigma2}
predicts zero string tension (as it should). When $N$
is even the string tension is given by 
\bea
\sigma = 2m\mu _{symm} \left ((1-\cos {4\pi \over N+2})+(N-2)(1-\cos {2\pi \over N+2})\right)
\eea

We discussed only the cases of the fundamental, adjoint,
anti-symmetric and symmetric representations since we used
bosonization techniques which are applicable to a limited class of
representations\cite{GNO}.

\appendix
\section{Appendix}
We give here a detailed derivation of the external quark anti-quark field
\eqref{external-charge}.

For the case of external charges in a real representation the $u$
field can chosen to point in some special direction in the $SU(N)$ algebra which we take to
be '3', namely $u=\exp -i T^3 \phi$. For external
charges in a complex representation one has to dress this ansatz with
a baryon number part, namely  $u=\exp -i\chi \exp
-iT^3\phi $. In both cases the gauge field is taken in the same
direction $A_+ = a_+ T^3$. The Lagrangian for the former case takes
the form

\beq
{\cal L} = {k\over 8\pi} (\partial _- \phi)( \partial _+ \phi ) +
{1\over 2e^2} (\partial _- a_+)^2 + M^2
\sum _i \cos \lambda _  i \phi + {k\over 4\pi} \partial _-\phi a_+,
\eeq
where $A_-=0$ gauge was used, $k$ is the level 
 and $\lambda _i$ are the isospin
components of the diagonal sub $SU(2)$ generator $T^3$.

The equations of motion for the matter and gauge fields are the
following
\bea
&& \label{phi} {k\over 4\pi} \partial _ - \partial _+ \phi + M^2 \sum _i \lambda _i
\sin \lambda _i \phi + {k\over 4\pi} \partial _- a_+=0 \\
&& \label{A} \partial _- ^2 a_+ = e^2 {k\over 4 \pi}  \partial _- \phi
\eea
Integrating \eqref{A} with zero boundary conditions and substituting in \eqref{phi} we obtain 
\beq
 {k\over 4\pi} \partial _ - \partial _+ \phi + M^2 \sum _i \lambda _i
\sin \lambda _i \phi + e^2 {({k\over 4\pi})}^2 \phi =0
\eeq
Let us assume a solution for $\phi$ which describes an infinitely
heavy light-cone static quark anti-quark system
\beq
\phi = \alpha \left (\theta (x^- +L)-\theta(x^- -L)\right ),
\eeq
where $\alpha$ is a yet unknown coefficient.

For the region $-L<x^-<L$ we obtain
\beq
\label{alpha}
M^2 \sum _i \lambda _i \sin \lambda _i \alpha +  e^2 {({k\over
    4\pi})}^2 \alpha =0
\eeq
When $M^2 \gg e^2$ the solution for $\alpha$ is of the following form 
\beq 
\alpha = 4\pi n + \epsilon,
\eeq
where $n$ is integer (we will pick the minimal $n=1$ possibility) and
$\epsilon$ is determined by the substitution in \eqref{alpha}
\beq
M^2 \sum _i \lambda _i ^2 \epsilon  +  e^2 {({k\over
    4\pi})}^2 4\pi \approx 0
\eeq  
 Thus $\alpha$ is given by
\beq 
\alpha = 4\pi - {e^2 \over M^2}  {({k\over
    4\pi})}^2 {4\pi \over \sum _i \lambda _i ^2} +
O\left ({({e^2\over M^2})}^2 \right)
\eeq
In the limit $M^2 \rightarrow \infty$, $u$ is 
\beq
u=\exp -i4\pi \left ( \theta (x^-+L )-\theta
  (x^--L )\right ) T^3
\eeq 
When $u$ is in a complex representation, namely $u$ is represented by 
$u=\exp -i\chi \exp -iT^3\phi$, we find by repeating the above
derivation the following expression
\bea
\lefteqn{u=}  \\ &&
\exp -i2\pi \left ( \theta (x^-+L )-\theta
  (x^--L )\right ) \exp -i4\pi \left ( \theta (x^-+L )-\theta
  (x^--L )\right ) T^3, \nonumber
\eea
for $U(N>2)$ and
\bea
\lefteqn{u=} \\ &&
\exp -i\pi \left ( \theta (x^-+L )-\theta
  (x^--L )\right ) \exp -i2\pi \left ( \theta (x^-+L )-\theta
  (x^--L )\right ) T^3, \nonumber
\eea
for $U(2)$.
Note that the $SU(2)$ part has a $2\pi$ prefactor.

\end{document}